\documentclass[useAMS,usenatbib,usegraphicx,usedcolumn]{mn2e}
\usepackage{txfonts}
\usepackage{lscape}
\newcommand\ion[2]{#1$\;${\small\rmfamily\@Roman{#2}}\relax}%
\newcommand{\pedix}[2]{\ensuremath{#1_{\,\mbox{\scriptsize #2}}}}
\newcommand{\apix}[2]{\ensuremath{#1^{\,\mbox{\scriptsize #2}}}}
\newcommand{\errUD}[2]{\ensuremath{^{+#1}_{-#2}}}
\newcommand{\cdof}{\ensuremath{C/\mbox{d.o.f.}}}
\newcommand{\dcdof}{\ensuremath{\Delta C/\mbox{d.o.f.}}}
\newcommand{\ie}{i.e.}
\newcommand{\eg}{e.g.}
\newcommand{\pgunobs}{PG~1700+518}
\newcommand{\nhgalunobs}{\ensuremath{2.26\times10^{20}}}
\newcommand{\redunobs}{\ensuremath{0.292}}
\newcommand{\posunobs}{RA$=$\apix{17}{h}\apix{01}{m}\apix{24.8}{s}, Dec$=$\apix{+51}{d}\apix{49}{m}\apix{20}{s}}
\newcommand{\mum}{\ensuremath{\,\mu\mbox{\scriptsize m}}}
\newcommand{\nh}{\ensuremath{\mbox{cm}^{-2}}}
\newcommand{\nhsym}{\ensuremath{N_{\mbox{\scriptsize H}}}}
\newcommand{\lum}{\ensuremath{\mbox{ergs~s}^{-1}}}
\newcommand{\flux}{\ensuremath{\mbox{ergs~cm}^{-2}\mbox{~s}^{-1}}}
\newcommand{\kev}{\ensuremath{\,\mbox{\scriptsize keV}}}
\newcommand{\ang}{\ensuremath{\,\mbox{\scriptsize \AA}}}
\newcommand{\ha}{\ensuremath{\mbox{H}\alpha}}
\newcommand{\hb}{\ensuremath{\mbox{H}\beta}}
\newcommand{\feka}{\ensuremath{\mbox{Fe~K}\alpha}}
\newcommand{\xmm}{{XMM-\emph{Newton}}}
\newcommand{\iras}{{\emph{IRAS}}}
\newcommand{\iso}{\emph{ISO}}
\newcommand{\spitzer}{{\emph{Spitzer}}}
\newcommand{\asca}{{\emph{ASCA}}}
\newcommand{\rosat}{{\emph{ROSAT}}}
\newcommand{\hst}{\emph{HST}}
\newcommand{\iue}{\emph{IUE}}

\title[\xmm\ first X-ray detection of the LoBAL quasar \pgunobs]{\xmm\ first X-ray detection of the LoBAL quasar \pgunobs}
\author[L.~Ballo et al.]{
  L.~Ballo$^{1}$,
  E.~Piconcelli$^{2}$, 
  C.~Vignali$^{3}$,
  and 
  N.~Schartel$^{4}$\\
  $^{1}$Instituto de F\'{\i}sica de Cantabria (CSIC-UC), 
  Avda. Los Castros s/n (Edif. Juan Jord\'a), E-39005 Santander  
  (Spain); ballo@ifca.unican.es \\
  $^{2}$Osservatorio Astronomico di Roma (INAF), via Frascati 33, 
  I-00040 Monteporzio Catone, Roma, (Italy) \\
  $^{3}$Dipartimento di Astronomia, Universit\`a degli Studi di Bologna, 
  via Ranzani 1, I-40127, Bologna, (Italy) \\
  $^{4}$\xmm\ Science Operation Centre, ESAC, ESA, P.O. Box 78, 
  E-28691 Villanueva de la Ca\ensuremath{\tilde{\mbox{n}}}ada, Madrid, (Spain)
}
\begin{document}

  \date{Accepted 2011 April 8. Received 2011 April 6; in original form 2011 March 4}

  \pagerange{\pageref{firstpage}--\pageref{lastpage}} \pubyear{XXXX}

  \maketitle

  \label{firstpage}

  \begin{abstract}
   We report the first high-energy detection of \pgunobs, a well-known low-ionization broad absorption line quasar
   (QSO).
   Due to previous X-ray non-detection, it was classified as soft X-ray weak QSO.
   We observed \pgunobs\ with \xmm\ for about $60\,$ksec divided in three exposures.
   The spectrum below $2\,$keV is very steep, $\Gamma\sim 2.4-3.8$, while at higher energies the extremely flat 
   emission (photon index $\Gamma\sim 0.15$, when
   modelled with a power law) suggests the presence of strong absorption ($\pedix{N}{H, pl}\sim2\times 10^{23}\,$\nh, 
   $\Gamma$ fixed to $1.8$), or a reflection-dominated continuum.
   The broad-band flux is consistent with previous non-detection.
   Simultaneous EPIC and OM data confirm its X-ray weakness ({\it observed} $\pedix{\alpha}{ox} \sim -2.2$). 
   The level of obscuration derived from the X-ray spectra of \pgunobs\ cannot explain its soft X-ray nuclear weakness
   unless a column density of $\pedix{N}{H}\gtrsim 2\times 10^{24}\,$\nh\ is 
   present.
   \end{abstract}

   \begin{keywords}
   galaxies: active -- quasars: individual: \pgunobs\ -- X-rays: galaxies
   \end{keywords}


\section{Introduction}\label{sect:intro}

The so-called ``soft X-ray weak QSOs'' (SXWQs, X-ray--to--optical spectral index
$\pedix{\alpha}{ox}\equiv \log (\pedix{F}{2\kev}/\pedix{F}{2500\ang})/\log (\pedix{\nu}{2\kev}/\pedix{\nu}{2500\ang}) \lesssim -2$, 
\citealt{laor97}; $\sim 10$\% of
nearby PG~QSOs, \citealt{brandt00}) are active galactic nuclei (AGN) notably 
faint in X-rays relative to their optical/UV fluxes.
Recently, \citet{gibson08}, studying the frequency of intrinsically SXWQs
in optically selected samples, found that [excluding Broad Absorption Line (BAL)] $\lesssim 2$\% SDSS QSOs are genuinely
X-ray weak.

Several possible explanations of this weakness have been proposed over the past years, all related to the X-ray band 
(the optical-UV flux in the SXWQs is not strongly variable and, therefore,
it is difficult to ascribe the X-ray weakness to a temporary high emission state at these wavelengths).
The presence of absorbers (neutral or partially ionized) can reduce in a considerable way the
observed high-energy emission 
(\citealt{gallagher01,gallagher05,piconcelli05}; see \eg\ Q1246-057 and SBS1542+541, 
\citealt{grupe03}; Mrk~304, \citealt{piconcelli04}; 
WPVS~007, \citealt{grupe07,grupe08}; Mrk~335, \citealt{grupeMrk08}; PG~1535+547, \citealt{ballo08}).
Strong UV absorption features of SXWQs classified as BAL sources are known to be related to warm absorbers
\citep{crenshaw03}.
Significant correlations have also been found between the degree of X-ray weakness \citep[\ie, how the
\pedix{L}{X} differ from that expected for a typical QSO having the same \pedix{L}{UV}; \eg,][and references therein]
{gibson08} and acceleration-dependent BAL properties in the BAL QSOs belonging to the SDSS
\citep{gibson09}.
A strong variability of the intrinsic X-ray continuum can be another reason of weakness at high
energies.
In this case, the classification as SXWQ is based on observations performed
during a low emission state of the source (see \eg\ PG~0844+349,
\citealt{gallo11}, and references therein;
1H~0419-577, \citealt{pounds04}; Mrk~335, \citealt{grupeMrk07}).
In this scenario, a reflection-dominated continuum originating in the accretion disc (caused by strong
light bending) can also play an important role
\citep[see][]{schartel07,schartel10,gallo11}.
Finally, the observed weakness can be an intrinsic property, due to a fundamentally distinct
emission mechanism (\citealt{risaliti03}; see also the case of PG~1254+047, \citealt{sabra01};
PHL~1811, \citealt{leighly07i}).
This last hypothesis implies different scenarios for the accretion disc.
Higher accretion rates can produce a steeper \pedix{\alpha}{ox} 
if the X-ray photons are emitted in a more central region than the UV-optical photons \citep{zdziarski04}, or
via the so-called ``photon-trapping'' \citep{begelman78,kawaguchi03}.
Another possibility is that the corona is radiatively inefficient \citep[\eg,][]{bechtold03,proga05}.
The most likely explanation for the dramatic steepening of \pedix{\alpha}{ox} reported by \citet{miniutti09} for
PHL~1092 seems to be a transient dramatic weakening or disruption of the X-ray corona.

\pgunobs\ \citep[$z=\redunobs$;][]{red} is the most luminous radio-quiet QSO in the optically selected PG 
sample \citep[absolute magnitude $\pedix{M}{B}=-25.2$;][]{veron10}.
The source resides in a system composed by two apparently interacting galaxies, with widely separated 
nuclei \citep[$\sim1.6\arcsec$, corresponding to $\sim7\,$kpc at the source redshift; \eg,][]{guyon06}.
Observational evidences are consistent with the companion being a collisional ring galaxy, possibly produced by a
near head-on collision with the QSO host.
The IR luminosity $\pedix{L}{8-1000\mum}\sim 5.5\times 10^{12}\,\pedix{L}{\sun}$
\citep[estimated from \iras, \iso, and \spitzer\ data;][]{evans09} 
implies for the whole system a classification as Ultra-Luminous Infrared Galaxy.
A rough indication for the bolometric luminosity, $\pedix{L}{bol}\sim 2.7\times 10^{46}\,$\lum,
was obtained by \citet{evans09} from this \pedix{L}{IR} 
\citep[adopting a ratio $\pedix{L}{IR}/\pedix{L}{0.1-1\mum}=0.76$;][]{guyon06}; given the black hole mass of
$\pedix{M}{BH}=7.81\times 10^{8}\,\pedix{M}{\sun}$ \citep{peterson04}, this implies a quite large Eddington ratios,
$\pedix{L}{bol}/(1.3\times 10^{38}\pedix{M}{BH})\sim 0.27$, as expected for this kind of sources.
Assuming that the luminosity in the far-IR corresponds to the luminosity of the starburst (SB), a star formation rate 
for the QSO host of $\mbox{\it SFR}\sim 210\,\pedix{M}{\sun}\,$yr$^{-1}$ is estimated \citep{evans09}.
Millimetre-wave CO(1$\rightarrow$0) observations allow the authors to derive for the system a molecular gas 
mass of $\pedix{M}{\pedix{H}{2}}\sim 5.7\times 10^{10}\,\pedix{M}{\sun}$ (mainly associated with the QSO host),
converting it into one of the most molecular gas-rich PG QSO host systems.
Moreover, \pgunobs\ belongs to the rare class of low-ionization broad absorption line (loBAL) QSOs, a very rare 
subclass ($\sim10$\%) of BAL QSOs showing absorption from both low- and high-ionization species.

At high energies, \rosat\ and \asca\ non-detections are the only X-ray information collected so far
\citep{green96,gallagher99,brandt00,george00}.
A power-law (PL) model with $\Gamma=2$ implies an \asca\ $\pedix{F}{2-10\kev}<8\times 10^{-14}\,$\flux\
\citep{george00}.
From the $3\sigma$ upper limit to the \rosat\ rate ($<3.4\times 10^{-3}\,$counts~s$^{-1}$, up to $\sim 2.5\,$keV), 
\citet{green96} estimated an index $\pedix{\alpha}{ox}<-2.30$.
With optical spectropolarimetric studies of the broad Balmer \ha\ and \hb\ emission lines, 
\citet{young07} demonstrate the presence of a wind launched from a relatively narrow annulus within
the accretion disc of \pgunobs.
The two-zones wind model adopted by the authors to interpret the observations can explain observed
properties such as BALs and X-ray absorbers.
Indeed, it requires the presence of an inner self-shielding region, opaque to X-rays.
On the other hand, the BALs should be formed in a part of the wind different from the X-ray-absorbing zone,
arising in the shielded outer wind.
This wind can be responsible for the extinction of an intrinsically normal X-ray emission.

In this paper we present the analysis of our \xmm\ observation of \pgunobs, providing the first X-ray
detection of this SXWQ.
In our analysis, we assume a cosmology with \pedix{\Omega}{M}$=0.3$, \pedix{\Omega}{$\Lambda$}$=0.7$, and
\pedix{H}{0}=$70\,$km~s$^{-1}$~Mpc$^{-1}$, implying a luminosity distance of $\pedix{D}{L} \approx
1505\,$Mpc.
All literature measurements have been converted to this cosmology.


\section{Observation and data reduction}

\xmm\ \citep{xmm} observed \pgunobs\ (\posunobs) in December~2009/January~2010, for about 
$60\,$ksec divided in three exposures, the first separated by about two weeks from the others 
(Obs.~ID $0601870101$, $0601870201$, and $0601870301$; hereafter, $101$, $201$, and $301$, respectively).
The observations were performed with the European Photon Imaging Camera (EPIC; pn, \citealt{pn}; MOS, \citealt{mos}), 
the Optical Monitor \citep[OM;][]{om}
and the Reflection Grating Spectrometer; the source is not detected with the last instrument.
The three EPIC cameras were operating in full frame mode, with the thin filter applied.
OM performed observations with the UVW1 ($\pedix{\lambda}{eff}=2910\,$\AA) and UVM2
($\pedix{\lambda}{eff}=2310\,$\AA) filters in the optical light path; all OM exposures were performed in 
the default image mode.

\begin{figure}
 \centering
 \includegraphics[height=4.cm,width=0.23\textwidth]{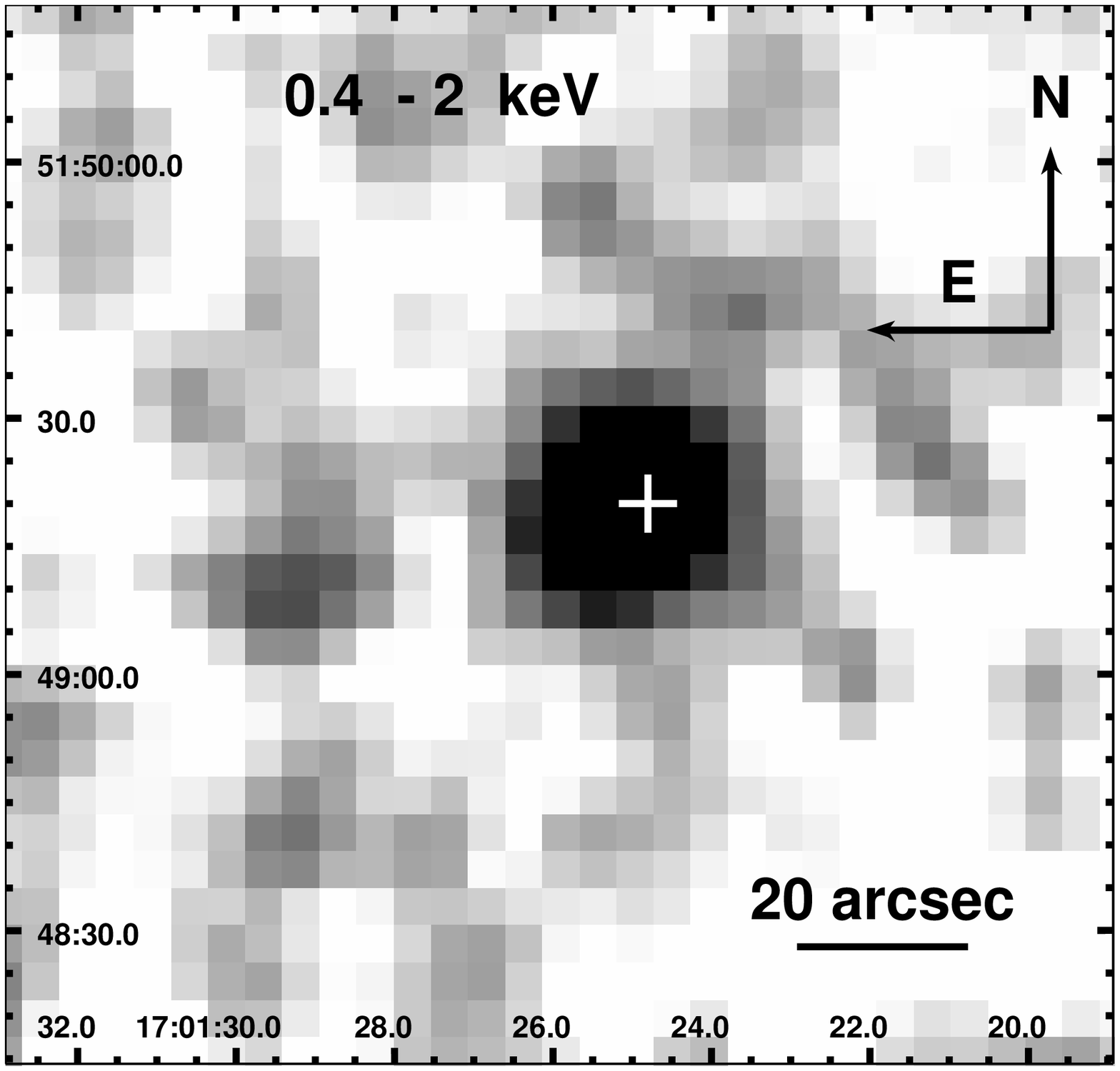}
 \hspace{0.1cm}
 \includegraphics[height=4.cm,width=0.23\textwidth]{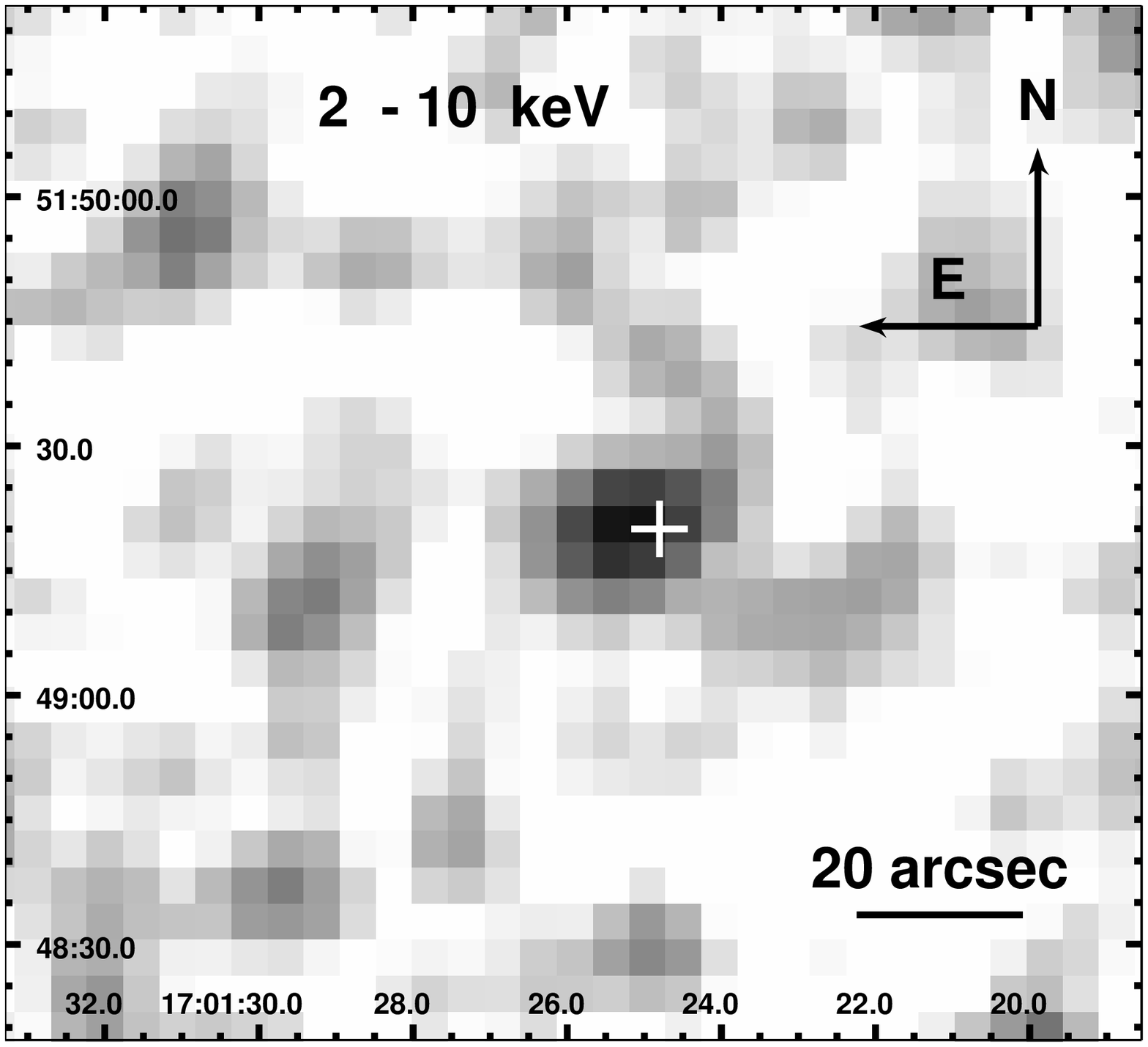}
 \caption{The $0.4-2\,$keV ({\it left}) and $2-10\,$keV ({\it right}) EPIC images (pn, MOS1, and MOS2 for
 observations $101$, $201$, and $301$ combined) of \pgunobs, smoothed with a Gaussian function with kernel 
 radius of $2$. 
 The white cross marks the optical position of the QSO.}
 \label{fig:image}%
 \vspace{-0.4cm}
\end{figure}

The data have been processed following the standard procedures, using the Science Analysis Software (SAS
version~10.0.2) with calibration from July~2010.
EPIC event files have been filtered for high-background time intervals, following the standard method
consisting in rejecting periods of high count rate at energies $>10\:\,$keV.
The net exposure times at the source position after data cleaning are $\sim9$, $16$, and $19\,$ks (MOS1),
$\sim15$, $17$, and $18\,$ks (MOS2), and $\sim4$, $9$, and $15\,$ks (pn).
Figure~\ref{fig:image} shows the EPIC images obtained combining event files from all cameras for the three 
exposures in the $0.4-2\,$keV ({\it left panel}) and $2-10\,$keV ({\it right panel}) energy ranges
(the white cross marks the optical position of \pgunobs).
The source is detected with a $0.5-2\,$keV EPIC detection likelihood above $5$ in all the observations
(according to the results of 
the SAS task {\it edetect\_chain} applied to the cleaned data); above $2\,$keV, there is
no detection in the observation $101$ (as well as in the MOS1 data of the $201$), while the source is marginally
detected in the three cameras during the observation $301$, and in the pn and MOS2 data of the $201$ (see
Table~\ref{tab:xmmlog}).

We extracted source counts from a circular region of $15\arcsec$ radius, enclosing 
both the QSO
host and the galaxy companion; background counts were extracted from a
source-free circular region in the same chip of $30\arcsec$ radius.
In Table~\ref{tab:xmmlog} we report the net count rates and 
signal-to-noise ratios ($S/N$) in the $0.4 - 2\,$keV and 
$2 - 10\,$keV observed energy ranges, corresponding to $0.5-2.6\,$keV and
$2.6-12.9\,$keV in the frame of the source.
No statistically significant variability has been detected during each observation. 
In Table~\ref{tab:xmmvar} we compare the observed variance, as evaluated from the data scatter,
to the variance expected from a constant source, as evaluated from the errors in the count rates.
This is done for background-subtracted data in bins of $500\,$sec each.
Given the relatively high black hole mass of \pgunobs, we note that even the longer exposure (after cleaning) is less 
than $3$ times the crossing time $\pedix{t}{cross}\equiv\pedix{R}{g}/c\sim 7700\,$sec.
Therefore, we do not expect variability in a single observation.

%
%

%
\begin{table*}
\begin{minipage}[!ht]{2\columnwidth}
\caption{{\bf EPIC {\xmm} data observation log.}}   
\label{tab:xmmlog}      
\begin{center}          
\renewcommand{\footnoterule}{}  
{\footnotesize
\begin{tabular}{c@{\extracolsep{0.3cm}} c@{\extracolsep{0.3cm}} c@{\extracolsep{0.3cm}} c c@{\extracolsep{0.4cm}} c c@{\extracolsep{0.3cm}} c }
\hline\hline       
   Obs. ID    & Instr. & Net exp. time & \multicolumn{2}{c}{Net Count Rates} & \multicolumn{2}{c}{$S/N$} & $(\pedix{H}{Rate}-\pedix{S}{Rate})$  \\
   \cline{8-8}
              &        & [sec]         & \multicolumn{2}{c}{[$10^{-3}\,$counts/sec]} & & & $(\pedix{H}{Rate}+\pedix{S}{Rate})$ \\
   \cline{4-5} \cline{6-7}
              &        &               & $S=0.4-2\,$keV & $H=2-10\,$keV & $S$ & $H$ &  \\
   \hline                  
   	  & MOS1  & $8816 $ & $1.7\pm0.5$ & $< 0.16$	& $3.47$ & $< 1.02$ & $< -0.82$      \\
   101    & MOS2  & $14520$ & $0.9\pm0.4$ & $< 0.47$	& $2.60$ & $< 2.76$ & $< -0.29$      \\
   	  & pn    & $4494 $ & $4.3\pm1.1$ & $< 1.6$	& $3.86$ & $< 2.11$ & $< -0.45$      \\  
   \hline  
   	  & MOS1  & $15630$ & $1.1\pm0.3$ & $< 0.28$	& $3.63$ & $< 1.50$ & $< -0.56$      \\
   201    & MOS2  & $16520$ & $1.1\pm0.3$ & $0.4\pm0.2$ & $3.58$ & $1.67$   & $-0.49\pm0.27$ \\
   	  & pn    & $9402 $ & $5.8\pm0.9$ & $1.1\pm0.6$ & $6.62$ & $2.12$   & $-0.67\pm0.14$ \\
   \hline
   	  & MOS1  & $18570$ & $1.2\pm0.3$ & $0.8\pm0.2$ & $4.47$ & $3.20$   & $-0.23\pm0.29$ \\
   301    & MOS2  & $17590$ & $0.9\pm0.3$ & $0.5\pm0.2$ & $3.49$ & $2.34$   & $-0.34\pm0.24$ \\
   	  & pn    & $15100$ & $4.4\pm0.6$ & $0.8\pm0.4$ & $7.45$ & $2.21$   & $-0.68\pm0.13$ \\
\end{tabular}
}
\end{center}          
\end{minipage}
\end{table*}
%

%
\begin{table}
\begin{minipage}[!ht]{\columnwidth}
\caption{{\bf Check for variability in the EPIC {\xmm} data.}}   
\label{tab:xmmvar}      
\begin{center}          
\renewcommand{\footnoterule}{}  
{\footnotesize
\begin{tabular}{c@{\extracolsep{0.0cm}} c@{\extracolsep{0.2cm}} r@{\hspace{0.2cm}/} l@{\extracolsep{0.4cm}} r@{\hspace{0.2cm}/\extracolsep{0.2cm}} l}
\hline\hline       
   Obs. ID    & Instr. & \multicolumn{4}{c}{Obs.\hspace{0.2cm}/\hspace{0.2cm}Exp. Variance} \\
              &        & \multicolumn{4}{c}{[$10^{-5}\,$(counts/sec)$^{2}$]} \\
   \cline{3-6}
              &        & \multicolumn{2}{c}{$0.4-2\,$keV} & \multicolumn{2}{c}{$2-10\,$keV} \\
   \hline                  
   	  & MOS1  & $0.71 \pm 0.16$ & $0.88 \pm 0.21$ & $0.43 \pm 0.10$ & $0.66 \pm 0.15$ \\
   101    & MOS2  & $0.64 \pm 0.15$ & $0.61 \pm 0.14$ & $0.24 \pm 0.06$ & $0.56 \pm 0.13$ \\
   	  & pn    & $6.42 \pm 1.50$ & $8.07 \pm 1.90$ & $6.43 \pm 1.50$ & $9.96 \pm 2.40$ \\  
   \hline  
   	  & MOS1  & $0.41 \pm 0.09$ & $0.43 \pm 0.10$ & $0.17 \pm 0.04$ & $0.30 \pm 0.07$ \\
   201    & MOS2  & $0.22 \pm 0.05$ & $0.44 \pm 0.10$ & $0.19 \pm 0.04$ & $0.34 \pm 0.08$ \\
   	  & pn    & $5.48 \pm 1.30$ & $8.08 \pm 1.90$ & $5.68 \pm 1.30$ & $8.67 \pm 2.00$ \\
   \hline
   	  & MOS1  & $0.33 \pm 0.07$ & $0.32 \pm 0.07$ & $0.25 \pm 0.06$ & $0.28 \pm 0.06$ \\
   301    & MOS2  & $0.28 \pm 0.06$ & $0.34 \pm 0.08$ & $0.16 \pm 0.04$ & $0.17 \pm 0.04$ \\
   	  & pn    & $2.10 \pm 0.50$ & $1.82 \pm 0.43$ & $0.90 \pm 0.21$ & $2.12 \pm 0.50$ \\
\end{tabular}
}
\end{center}          
\end{minipage}
\end{table}

For each observation, we obtained two images for each OM filter. 
The count rate information was averaged;
the resulting AB magnitudes are: in the UVM2 filter, $16.01\pm0.01$, $16.00\pm0.01$, and $16.00\pm0.01$; in the UVW1
filter, $15.70\pm0.01$, $15.67\pm0.01$, and $16.68\pm0.01$ (for observations $101$, $201$, and $301$,
respectively).


\section{X-ray spectral analysis}

EPIC spectra have been analysed using standard software packages 
\citep[FTOOLS version~6.10; XSPEC version~12.6.0q,][]{xspec}.
All the models discussed assume Galactic absorption with a column density of $\pedix{N}{H,
Gal}=\nhgalunobs\:\,$\nh\ \citep{nh}.
Unless otherwise stated, the quoted luminosities are corrected for both Galactic and intrinsic absorption. 
Because of the small number of counts in our spectra, we decided
to use the Cash-statistic \citep[{\sc cstat} command in XSPEC, now allowing fits to data for which 
background spectra are considered]{cash79}, binning the spectra to have at least $1$ count in each 
channel\footnote{See https://astrophysics.gsfc.nasa.gov/XSPECwiki/low\_count\_spectra, section ``Use a modified cstat''}.

We created a combined MOS spectrum and response matrix for each observation, to improve the $S/N$ ratio.
The $301$, $201$, and $101$ spectra present the same
spectral shape and intensity (see the hardness ratios in the last column of Table~\ref{tab:xmmlog}).
We therefore combined the pn [MOS] data, fitting the spectra simultaneously
between $0.4$ and $10\,$keV (for a total of $296$ bin); the relative 
normalizations, left free during the fit, are consistent within a few percent.

\begin{figure}
 \centering
 \resizebox{0.95\hsize}{!}{\includegraphics[angle=270]{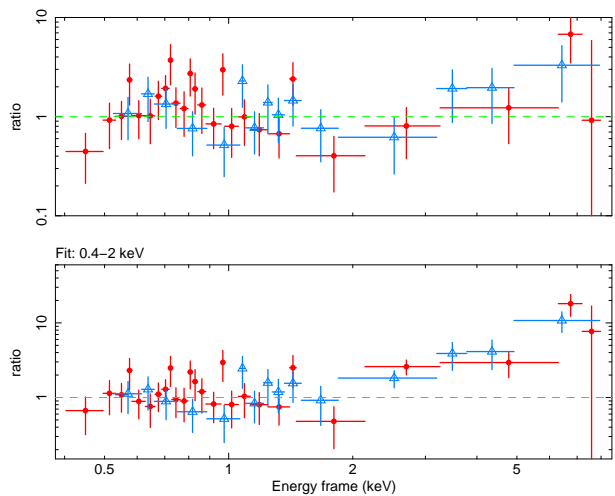}}
 \caption{Data-to-model ratio for the pn (red filled
 circles) and MOS (blue open triangles) data (observations $101$, $201$, and $301$ combined).
 The adopted models are a simple power law fitted over the whole range ({\it upper panel}),
 and an absorbed power law plus a {\sc mekal} component (with fixed intensity, see text) fitted in the 
 $0.4-2\,$keV, and then extrapolated up to $10\,$keV ({\it lower panel}).
 For graphical purposes, data have been binned to a $2\sigma$ significance.}
 \label{fig:pl}%
\end{figure}

A simple power-law model plus Galactic absorption
($\Gamma=2.1\pm 0.3$, $C=281.7$ for $293$ d.o.f.) leaves positive residuals both below $1\,$keV 
(more pronounced in the pn than in the MOS data) 
and above $4\,$keV, while a trough characterizes the emission between $1.5$ and $3\,$keV (see
Fig.~\ref{fig:pl}, {\it upper panel}).
This simple model implies a rather low luminosity, $\pedix{L}{2-10\kev} = (2.5\pm 0.4)\times 10^{42}\,$\lum.

%

Limiting to $E<2\,$keV, where most of the flux is emitted, confirms this result: the 
spectra are well described ($\cdof=147.2/176$ for $180$ bins) by an absorbed power-law model with a photon 
index $\Gamma=3.8\errUD{1.2}{1.0}$ and a 
column density $\nhsym=3.3\errUD{2.4}{2.0}\times 10^{21}\,$\nh; the addition of the intrinsic 
absorption implies a change in the statistics $\Delta C$=$7.5$ for $1$ d.o.f. less, 
with respect to a single power law ($\Gamma=2.2\pm 0.3$).
In low-quality data, a steep power law
can mimic the large-scale scattered/reprocessed emission observed in obscured Seyfert~2
spectra \citep[see \eg][]{guainazzi07}, characterized (in higher resolution data) by soft X-ray emission lines, 
and likely associated with the Narrow Line Region \citep{bianchi06}.

On the other hand, in the soft X-ray band we expect a thermal contribution associated with the strong SB observed in 
this source.
Fitting the data below $2\,$keV with a {\sc mekal} model (describing in XSPEC a collisionally 
ionized plasma emission component) with H density fixed to $1\,\mbox{cm}^{-3}$ and metal abundance fixed to the 
solar one, we obtain an implausibly high temperature
($kT = 2.6\errUD{2.1}{0.9}\,$keV; $\cdof=163.3/177$); a more typical temperature, $kT = 0.6\pm 0.1\,$keV,
is obtained with the addition of intrinsic absorption, $\nhsym=7.2\errUD{0.7}{1.5}\times 10^{21}\,$\nh\ 
($\Delta C$=$13.5$ for $1$ d.o.f. less).
Left free to vary, the metal abundance is rather low and poorly constrained, $Z=0.05\errUD{0.12}{0.04}$, with a
statistical upper limit consistent with typical results for SB galaxies \citep[$Z<0.5$, \eg,][]{ptak99}. 

The soft X-ray luminosity associated to this component,
$\pedix{L}{0.4-2\kev, MEKAL} = 2.1\errUD{1.6}{0.8}\times 10^{43}\,$\lum, is about $20$ times higher than expected 
from a $\mbox{\it SFR}\sim 210\,\pedix{M}{\sun}\,$yr$^{-1}$, (as estimated by 
\citealt{evans09} assuming that \pedix{L}{FIR} is the SB luminosity).
According to \citet{ranalli03}, the soft X-ray luminosity corresponding to this {\it SFR} is
indeed $\pedix{L}{0.5-2\kev, SFR}\sim 10^{42}\,$\lum.
A contribution from the AGN is likely present at the mid-IR/FIR wavelengths which are used 
in the estimate of the {\it SFR}; therefore, this value can be assumed as
an upper limit to the soft X-ray luminosity expected from the SB.
The present data do not allow to disentangle the SB-related from the AGN-related X-ray emission
leaving free to vary all the parameters of interest; 
however, we can investigate the relative contribution of AGN and SB by fixing the intensity of the latter to the 
expected upper limit.
Residuals are well modelled by a power law flatter than previously found but still consistent, given the large 
associated uncertainties, $\Gamma=2.4\errUD{1.6}{1.1}$, with a luminosity
$\pedix{L}{0.4-2\kev, PL} = 3.1\errUD{5.7}{1.7}\times 10^{42}\,$\lum; the parameters are poorly constrained, with
$\nhsym=1.2\errUD{2.5}{1.0}\times 10^{21}\,$\nh\ and $kT = 0.7\errUD{0.4}{0.3}\,$keV ($\cdof=141.3/175$).

When extended to the whole energy range, strong residuals at high energy are clearly evident (see
Fig.~\ref{fig:pl}, {\it lower panel}).
A simple power law fitted to the data above $2\,$keV results to be extremely flat,
$\Gamma=0.2\errUD{0.8}{0.9}$, leaving a strong soft excess below $2\,$keV.

Such an extremely flat power law suggests the presence of absorption, or it can be interpreted as a reflection 
component.
We tested both hypotheses (\ie, transmission vs. reflection-dominated scenarios), adopting for the latter the 
{\sc pexrav} model in XSPEC \citep{pexrav}.
The statistic quality of the data above $\sim3\,$keV is too low to constrain at the same time all the spectral 
parameters for the reflection component or the high-energy absorbed power law; in the following, we fix the photon index
of the intrinsic nuclear emission to a typical value for unabsorbed AGN, $\Gamma=1.8$
\citep[\eg,][]{piconcelli05,mateos10}.
Both the adopted models are adequate to describe the overall broad-band shape, with values for the common model
parameter, the soft-component photon index, consistent within the errors: $\pedix{\Gamma}{soft}=4.5\errUD{1.3}{1.1}$
and $3.9\errUD{1.3}{0.9}$, when a reflection-dominated continuum or a transmission scenario are assumed, 
respectively. 
A fit with an absorbed power law implies a high column density of 
$\pedix{N}{H, pl}=2.4\errUD{4.0}{1.7}\times 10^{23}\,$\nh.
Furthermore, an additional absorption of $\pedix{N}{H}=4.2\errUD{2.5}{2.1}\times 10^{21}\,$\nh\ and 
$3.4\errUD{2.4}{1.9}\times 10^{21}\,$\nh, depending on the continuum model, covering both spectral components, is also 
required.
From a statistical point of view, at high energy a transmission or a reflection-dominated model are 
indistinguishable ($C=262.5$ for $290$ d.o.f. vs. $262.2$ for $291$ d.o.f.); the low $S/N$ prevents us from completely 
ruling out any possibility.

When the expected soft contribution from SB is included in the model, $\pedix{L}{0.5-2\kev, MEKAL}\sim
10^{42}\,$\lum, we found a flatter soft AGN-related power law, $\pedix{\Gamma}{soft}=1.9\pm 0.5$ and
$2.0\pm 0.6$ for the transmission or the reflection-dominated scenario, respectively.
In the former case, the absorber covering the intrinsic nuclear power law has a column density
$\pedix{N}{H, pl}=6.1\errUD{51.9}{1.5}\times 10^{23}\,$\nh.
The additional absorption covering the all the spectral components is no longer required ($\dcdof$=$2.0/1$ and 
$3.0/1$); the two scenarios are still indistinguishable ($C=259.5$ for $290$ d.o.f. vs. $C=261.6$ for $291$ d.o.f.).

\begin{figure}
 \centering
 \resizebox{0.95\hsize}{!}{\includegraphics[angle=270]{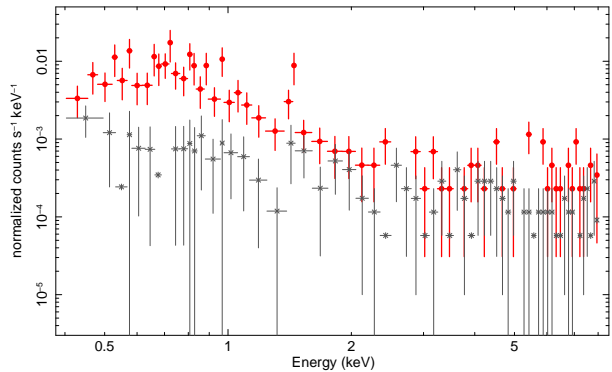}}
 \caption{Source plus background pn spectrum (red filled circles), for the three observations
 combined, compared with the background (black crosses).
 For graphical purposes, data have been binned to a $2\sigma$ significance.}
 \label{fig:lda}%
\end{figure}

In principle, 
the equivalent width of the \feka\ line can help in discriminating between a transmission or a reflection-dominated
origin of the high-energy emission, but this feature is not detected.
No meaningful upper limit for the line equivalent width can be obtained, due to the high background, dominating above 
$2\,$keV (see Fig.~\ref{fig:lda}).

\begin{figure}
 \centering
 \resizebox{0.95\hsize}{!}{\includegraphics[angle=270]{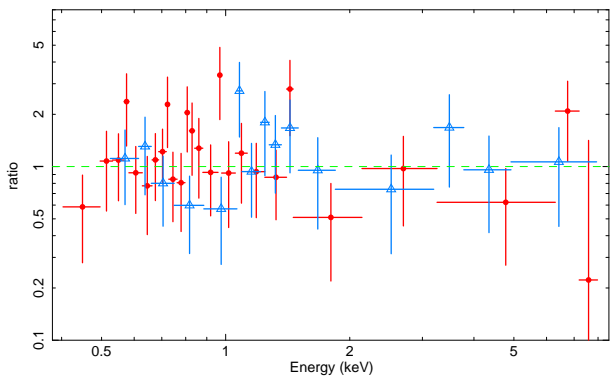}}
 \caption{Background-subtracted data-to-model ratio for the pn (red filled
 circles) and MOS (blue open triangles) for the transmission scenario.
 The fitting model consists of an absorbed power law, an unabsorbed power law,  
 and a {\sc mekal} thermal emission component (with fixed intensity, see text).
 For graphical purposes, data have been binned to a $2\sigma$ significance.}
 \label{fig:ra}%
\end{figure}

For an high-energy absorbed power-law model, we measure a total
$0.4-2\,$keV [$2-10\,$keV] flux (corrected for Galactic absorption) of $9.7\errUD{10.6}{4.5}\times 10^{-15}\,$\flux\ 
[$1.84\errUD{2.69}{0.99}\times 10^{-14}\,$\flux]. 
The total luminosity in the hard band is
$\pedix{L}{2-10\kev} = 1.0\errUD{1.5}{0.5}\times 10^{43}\,$\lum.
When the SB contribution is included, as a thermal component with $\pedix{L}{0.5-2\kev, MEKAL}\sim 10^{42}\,$\lum,
the hard X-ray luminosity associated with the obscured power-law continuum, representing the primary AGN emission in 
the transmission scenario, is $\pedix{L}{2-10\kev, PL} \sim 1.3\times 10^{43}\,$\lum.
Data-to-model ratio for this spectral fit is shown in Figure~\ref{fig:ra}.

Finally, the total observed $0.4-10\,$keV flux measured with \xmm, 
$2.72\errUD{3.65}{1.39}\times 10^{-14}\,$\flux, is consistent with the upper limit in the $0.6-10\,$keV 
\asca\ SIS0 counts reported by \citet{george00}, \ie\ $<3\times10^{-3}\,$counts~s$^{-1}$:
assuming the steep power law$+$absorbed power law model best fitting the \xmm\ data, 
{\sc WebSpec}\footnote{See http://heasarc.gsfc.nasa.gov/webspec/webspec.html} predicts an
\asca\ SIS0 count rate between $0.6$ and $9.5\,$keV of $\sim0.87\times10^{-3}\,$counts~s$^{-1}$.

%
%

%
%


\section{Discussion and conclusions}\label{sect:dis}

We have presented the first X-ray detection of the loBAL SXWQ \pgunobs, obtained from our \xmm\ observations.
Most of the flux is emitted below $2\,$keV; above this energy, the signal is weak
(see Fig.~\ref{fig:lda}).
The observed flux is consistent with previous \asca\ non-detection in this source.

The soft emission is well described by a steep power law with $\Gamma\sim 3.8$, 
or assuming a thermal component with $kT\sim0.6\,$keV.
However, the luminosity derived for this component is a factor of $20$ higher than expected from the {\it SFR}, and 
suggests that the emission is mainly due to reprocessing of the AGN emission \citep[\eg,][]{guainazzi07}, 
with a contribution due to the SB activity.
When the thermal emission is rescaled to the upper limit estimated on the basis of the {\it SFR},
the lower limit to the AGN contribution 
to the soft X-ray emission, represented by a power law with $\Gamma\sim 2.4$, is
$\pedix{L}{0.4-2\kev, PL}\sim 3\times 10^{42}\,$\lum.
Above $2\,$keV, the extremely flat power law ascribed to the AGN can be due to both an high-absorbed power law 
($\pedix{N}{H, pl}\sim2\times 10^{23}\,$\nh, $\Gamma$
fixed to $1.8$) in a transmission scenario, or to a reflection-dominated continuum.
While a strong absorption is not unexpected due to the loBAL nature of \pgunobs, 
the quality of the data is too low to discriminate between these possibilities, or to constrain more 
physical models: \eg, to test the ionization state of the absorbing gas (taking into account the BAL nature of this
QSO), or to investigate where the hypothetical reflection could take place: distant obscuring matter,
implying a Compton-thick nature, or the accretion disk, possibly contributing to the observed soft excess 
(as in the case of PG~2112+059, \citealt{schartel07,schartel10}, or PG~0844+349, \citealt{gallo11}).

No conclusive result is obtained considering the AGN-dominated mid-IR continuum,
that can provide an absorption-independent measure of the intrinsic luminosity even for Compton-thick sources.
Assuming the $\pedix{L}{5-6\mum}\sim 1.2\times 10^{45}\,$\lum\ reported by \citet{shi07}, the
absorption-corrected $\pedix{L}{2-10\kev}\sim 1.3\times 10^{43}\,$\lum\ for the power-law component would be consistent 
(taking into account the strong uncertainties) with both Compton-thin and Compton-thick nature 
\citep[see \eg\ fig.~4 in][]{alexander08}.
Note that, at face value, from the low luminosity obtained considering a simple power-law model, 
$\pedix{L}{2-10\kev}\sim 2.5\times 10^{42}\,$\lum, \pgunobs\ would lie in the region 
where Compton-thick sources are expected, in the same diagnostic diagram, thus 
reinforcing the hypothesis of strong obscuration.
Insight into AGN energetics can be obtained from mid-IR emission-line diagnostics.
The low ratio between high- and low-ionization line observed in \pgunobs,
$\mbox{[\ion{Ne}{v}]}/\mbox{[\ion{Ne}{ii}]}\leqslant0.2$ \citep{evans09}, typical of energetically weak AGN,
is similar to the value observed in the loBAL QSO Mrk~231 ($\mbox{[\ion{Ne}{v}]}/\mbox{[\ion{Ne}{ii}]}\leqslant0.15$, 
\citealt{armus07}), where a Compton-thick absorber \citep{braito04} is able to block the majority of the
$\mbox{[\ion{Ne}{v}]}$ line.
Non-detection of polycyclic aromatic hydrocarbon (PAH) features in the \spitzer\ IRS spectrum
\citep{shi07} of \pgunobs\ can be due to a combination of a mid-IR AGN emission partially sweeping away the PAH, and a
not strong contribution from star formation activity: note that Mrk~231, where 
aromatic features, although weak, are observed \citep{armus07}, has a 
$\mbox{\it SFR}\sim 470\,\pedix{M}{\sun}\,$yr$^{-1}$ \citep{farrah03}, twice the value found for
\pgunobs.
Summarizing, mid-IR data are not able to distinguish between an absorbed AGN and a Compton-thick AGN scenarios.

%
\begin{figure*}
  \centering
   \resizebox{0.95\hsize}{!}{\includegraphics[angle=90,width=4cm,height=2.3cm]{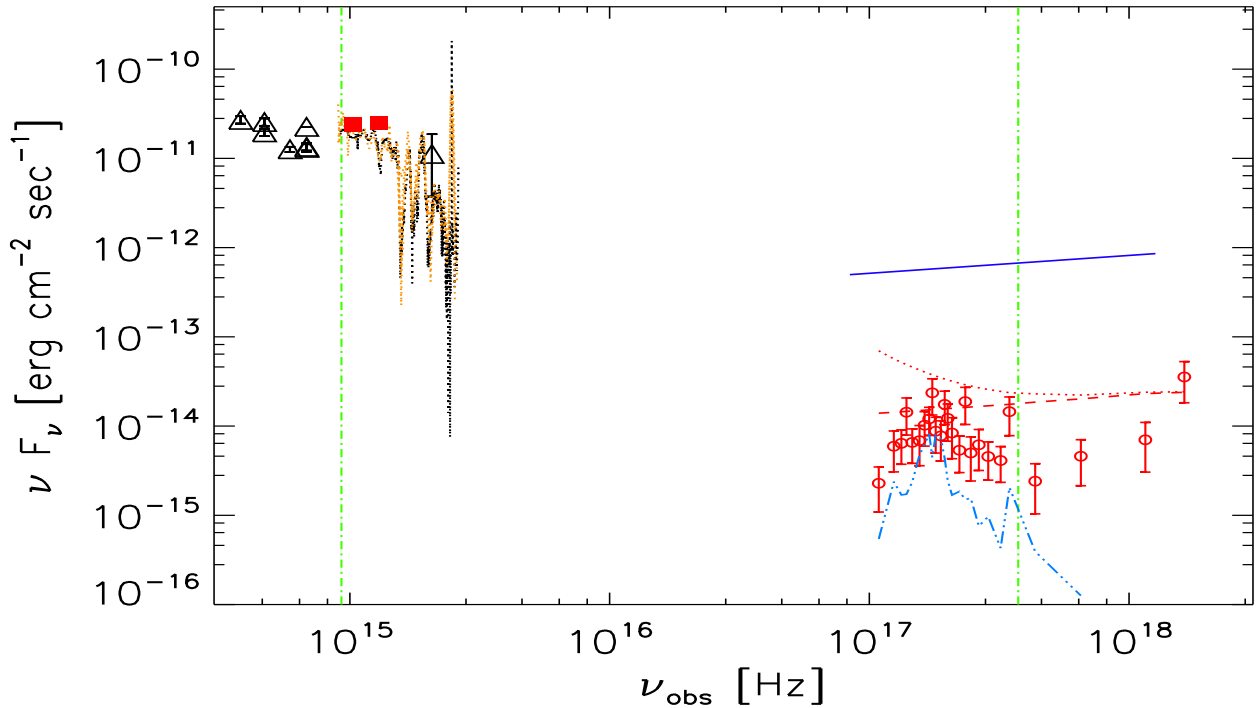}}
   \vspace{0.2cm}
  \caption{Spectral energy distribution for \pgunobs.
	   The vertical lines mark $\pedix{\lambda}{rf}=2500\,$\AA\ and $\pedix{E}{rf}=2\,$keV.
           \xmm\ data: filled squares, OM; open circles, observed \xmm\ pn flux, 
	   binned to a $2\sigma$ significance for graphical purposes.
	   Red dotted and dashed lines represent the whole emission and the hard power-law component alone, 
	   both unabsorbed, while the light blue dot-dot-dashed curve represents the expected SB emission.
	   The blue solid line indicates the X-ray flux (power law with $\Gamma=1.8$) expected
	   from OM data assuming an $\pedix{\alpha}{ox}=-1.59$,
	   typical for QSOs with the \pedix{L}{UV} of \pgunobs\ \citep{gibson08}.
	   At lower energies, we show \hst-FOS (black dotted line) and \iue\ (yellow dotted line) spectra.
	   Open triangles are optical data from NED.
  }
  \label{fig:multinused}%
\end{figure*}

We can use the OM observations with the UV filters to characterize the low-energy 
emission and to construct a broad-band spectral energy distribution (SED) of \pgunobs\ from {\it simultaneous} UV and
X-ray data (in principle, variability can change the classification as SXWQ).
The OM flux is compatible, within the errors, with previous observations: in 
Figure~\ref{fig:multinused} the OM data (red squares) are overplotted to 
\hst-FOS\footnote{See http://archive.stsci.edu/.} and 
\iue\footnote{See http://sdc.laeff.inta.es/ines/index2.html} spectra (dotted lines).
At higher energies, the EPIC pn data (red open circles) are plotted.
From the OM and EPIC data 
\citep[corrected for Galactic absorption, following the extinction curve provided by][]{cardelli89}, we compute 
an observed $\pedix{\alpha}{ox} = -2.19\pm 0.05$. 
From the relations between the broad-band spectral index and the optical luminosities 
obtained for 
samples of optically-selected AGN \citep[\eg,][and references therein]{gibson08}, we would expect from the 
$\pedix{L}{UV}$ of 
\pgunobs\ a value of $\pedix{\alpha}{ox} = -1.59\pm 0.16$ (where the uncertainty represents the
dispersion in the $\pedix{\alpha}{ox}-\pedix{L}{UV}$ relation); this would imply the X-ray emission marked
with the dotted line in Figure~\ref{fig:multinused}.
This supports the hypothesis of soft X-ray weakness for \pgunobs.

The observed \pedix{\alpha}{ox} is typical of optically-selected BAL QSOs
\citep[$\pedix{\alpha}{ox}\sim -2.5\;\>\mbox{-}\; -2$,][]{gallagher06,green96}, while it is more negative
than the values found by
\citet{giustini08} for a sample of X-ray selected BAL QSOs (consistent, within the errors, with the index 
expected from the correlation with $\pedix{L}{UV}$).
We obtain a low \pedix{\alpha}{ox} value, \ie\ $\pedix{\alpha}{ox} = -2.01\pm 0.20$,
even if we consider the absorption-corrected value of the AGN-related luminosity at $2\,$keV
(dashed line in Fig.~\ref{fig:multinused}). 
This means that, although the strong absorption observed in the source (typical of both SXWQs and
BAL QSOs) can be partially responsible of the observed low
\pedix{\alpha}{ox}, its effect is not enough to completely explain the soft X-ray weakness of the source.
A similar result was found for PG~1254+047, that is not only an intrinsically X-ray weak BAL QSO, 
like \pgunobs, but also
heavily X-ray absorbed, with a $\nhsym\gtrsim 2\times10^{23}\,$\nh\ ionized absorber \citep{sabra01}.
On the other hand, the column density required to obtain the expected $\pedix{\alpha}{ox} \sim -1.59$ (after correcting
for absorption) is consistent with a Compton-thick scenario, $\pedix{N}{H}\sim1.5\times 10^{24}\,$\nh.

\pgunobs\ is therefore a very peculiar source, thanks to the simultaneous presence of outflowing wind produced in the
innermost regions \citep{young07}, high nuclear obscuration, and strong star formation activity.
These properties  make this system similar to Mrk~231, for which \citet{feruglio10} provided
direct observational evidence of QSO-driven feedback on the host-galaxy gas content.
The host galaxy of \pgunobs\ is part of a collisionally interacting system, that is also one of the most 
molecular gas-rich PG QSO hosts observed to date.
This could suggests that also in \pgunobs\ we are observing an early dust-enshrouded phase in
the QSO evolution, predicted by the theoretical models of QSO activation by galaxy interactions \citep{dimatteo05}.
Deeper X-ray observations are needed, to properly describe the observed emission, to determine the physical
origin of the different components, and to weight the contribution of AGN and SB.
Moreover, information on possible variability in spectral shape and/or intensity is fundamental
to investigate the nature of the X-ray weakness of \pgunobs.

\section*{Acknowledgments}

  We warmly thank the referee, for her/his constructive comments that significantly improved the paper.  
  Based on observations obtained with \xmm\ (an ESA science mission with instruments and contributions
  directly funded by ESA Member States and the USA, NASA).
  LB acknowledges support from the Spanish Ministry of Science and Innovation through a ``Juan de la Cierva'' 
  fellowship.  
  Financial support for this work was provided by the Spanish Ministry of Science and Innovation, through 
  research grant AYA2009-08059.
  Support from the Italian Space Agency is acknowledged by EP (contract ASI/INAF I/088/06/0) and 
  CV (contracts I/088/06/0 and I/009/10/0).

\bibliographystyle{mn2e} 

\label{lastpage}

\end{document}